# COMMISSIONING OF IH-RFQ AND IH-DTL FOR THE GSI HIGH CURRENT LINAC

W. Barth, P. Forck, J. Glatz, W. Gutowski, G. Hutter, J. Klabunde, R. Schwedhelm,
P. Strehl, W. Vinzenz, D. Wilms, GSI Darmstadt, Germany
U. Ratzinger, Institute for Applied Physics, University Frankfurt am Main, Germany

*Abstract*

The new 1.4 MeV/u front end HSI (HochStromInjektor) of the Unilac accelerates ions with A/q ratios of up to 65 and with beam intensities in emA of up to 0.25 A/q. The maximum beam pulse power is up to 1300 kW. During the stepwise linac commissioning from April to September 1999 the beam behind of each cavity was analysed within two weeks. A very stable $Ar^{1+}$ beam out of a volume plasma source MUCIS was used mainly. The measured norm. 80 % emittance areas around 0.45 π mm mrad are close to the results from beam simulations. Up to 80 % of the design intensity at the linac exit were achieved. In February 2000 an $U^{4+}$ beam from the MEVVA source was accelerated for the first time.

## 1 INTRODUCTION

H-type RFQ- and DTL-structures are well suited and frequently used meanwhile at the front end of proton and heavy ion linacs. The recent GSI heavy ion linac project is characterized by three quite ambitious parameters: The high A/q-value of up to 65, the high beam current – for example 15 emA of $^{238}U^{4+}$, and the normalized horizontal emittance of less than 0.8 π mm mrad for the $U^{73+}$ beam at synchrotron injection after having passed two stripping processes. Multiturn injection into the horizontal phase space of the synchrotron SIS 18 should then allow the accumulation and acceleration of up to $4 \cdot 10^{10}$ $U^{73}$-ions. This aim as well as the state of the art in ion source development dictated the parameter choice for the new linac [1]. The original Wideröe DTL section was replaced by H-type structures (Fig. 1). They allow to increase the voltage gain by a factor of 2.5 while keeping the total length of the installations. Additionally, the injection energy into the Unilac was reduced from 11.4 keV/u to 2.2 keV/u. These changes allow a convenient matching of high current beams from MUCIS or MEVVA ion sources to the Unilac [2,3,4]. Several aspects of the new 91 MV linac are described in ref. [5,6,7]. The new 36 MHz power amplifier is described in ref. [8]. The choice of this frequency, which is one third of the succeeding Alvarez-DTL frequency resulted from beam dynamics calculations as well as from corresponding cavity sizes [7]. RF- and beam results from the IH-RFQ and from the Super Lens SL, which were invented during this project, are reported in this paper as well as the corresponding results of the IH-DTL.

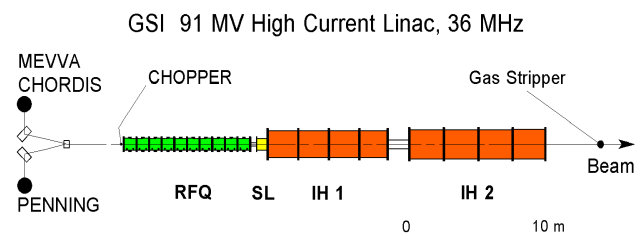

Fig. 1: General drawing of the HSI

## 2 TIME SCHEDULE

The main injector of the GSI complex had to be replaced at the end of this project. A nine month-period was permitted to remove the Wideröe-DTL as well as to install and to commission the new injector, rf equipment and the 1.4 MeV/u charge separator. During that period all GSI experiments were supplied exclusively by the High Charge State Injector [9]. The linac assembly and commissioning was done tank by tank and a beam diagnostics bench [10] was installed behind the component under investigation. The following time schedule shows the key events:

| | | |
|---|---|---|
| 17. Dec. | 1998 | Last beam from the Wideröe structure |
| March | 1999 | Beam injection into the 2.2 keV/u transport line |
| 28. April | 1999 | Acceleration by the IH-RFQ |
| 31. May | 1999 | Beam injection into the Super Lens |
| 22. July | 1999 | Acceleration by the IH-DTL, tank 1 |
| 06. Sept. | 1999 | Acceleration by the IH-DTL, tank 2 and beam transport to the gasstripper |

## 3 COMPONENT TESTS AND ALIGNMENT

The rf tuning results from all cavities as well as rf power tests with the first and the last of 10 RFQ modules are described in ref. [11]. The flatness of the voltage distribution along the mini vanes is within ± 1 %. The vane alignment within each module is given by the precision of the inidividual components. No adjustment during the final assembly was foreseen. The deviation in the transversal position of the vane-carrier rings was below 0.05 mm. The precision of each carrier ring is within

± 0.01 mm. The RFQ modules were then bolted together and after the final installation in the Unilac tunnel the vertical module displacement was up to 0.1 mm along a sine-shaped half wave, while the horizontal displacement was up to ± 0.25 mm along an S-shaped curve. As the total RFQ length is 9.4 m, the related angular deviations from the beam axis are quite acceptable.

The Super Lens [12], an 11-cell RFQ with enlarged aperture and synchronous phase – 90°, acts as a 3-dimensional focusing lens into the IH-DTL and is bolted on the low energy end plate of cavity IH1. The large tank modules of the DTL cavities were manufactured and copper plated in time as well as the bulk copper drift tubes. Some technical effort became necessary during the design and construction of the cavity internal quadrupole triplets: the cores are laminated and the material is cobalt steel alloy (Vacoflux50, Fa. Vacuumschmelze). The quadrupoles with core aperture diameters of 38 mm and 50 mm reach up to 1.27 T and 1.23 T, respectively, at the pole tips. These fields were demonstrated successfully in all cases. Due to the complex pulse structure of the Unilac the magnets have to allow pulsed operation with a pulse rise time of 16 ms only and pulse repetition rates of up to 20 Hz. This corresponds to voltages of up to 700 V applied on the coils. With the exception of one singlet, which is showing intolerable leak currents at repetition rates above 5 Hz, all lenses have reached the specifications. The present situation causes no restriction with respect to synchrotron injection. Anyway the corresponding triplet will be replaced at the first opportunity. The alignment results of the triplets are ± 0.1 mm for the magnetic axis of each singlet within the corresponding triplet. The deviations of all linac components from the beam axis are up to ± 0.25 mm. The triangulation measuring technique was applied.

After final installation in the Unilac tunnel the vacuum pressure became better than $2 \cdot 10^{-8}$ hPa for all structures within one month. All cavities came on rf power after about 2 days of preconditioning, where only some 10 W were accepted at vacuum pessures up to $7 \cdot 10^{-6}$ hPa. The RFQ as well as the Super Lens show pronounced dark current contributions at voltage amplitudes above the 75 % design level [7]. IH1 shows modest dark current contributions while IH2 seems to become free of that effect after some more operation time. $^{238}U^{4+}$-levels (that is 90 % of the A/q = 65 design level) were provided and used in beam times since 02/2000.

## 4 BEAM COMMISSIONING

The mobile diagnostics instrumentation consisted of 4-segmented capacitive pick-up probes, beam transformers, slit-grid as well as single shot, pepper pot type emittance measurement devices and diamond detectors for bunch length detection [10].

*Beam Energy:*
The time of flight technique was used to measure the beam energy [5-7]. In all cases the correct energy was received immediately after first beam injection and at the nominal rf parameters.

Additionally, the exit energy dependence on rf amplitude and phase can be measured and compared to beam simulations. Fig. 2 shows as an example the energy profile of IH2, which was calculated by LORASR and verified quite well by TOF measurements.

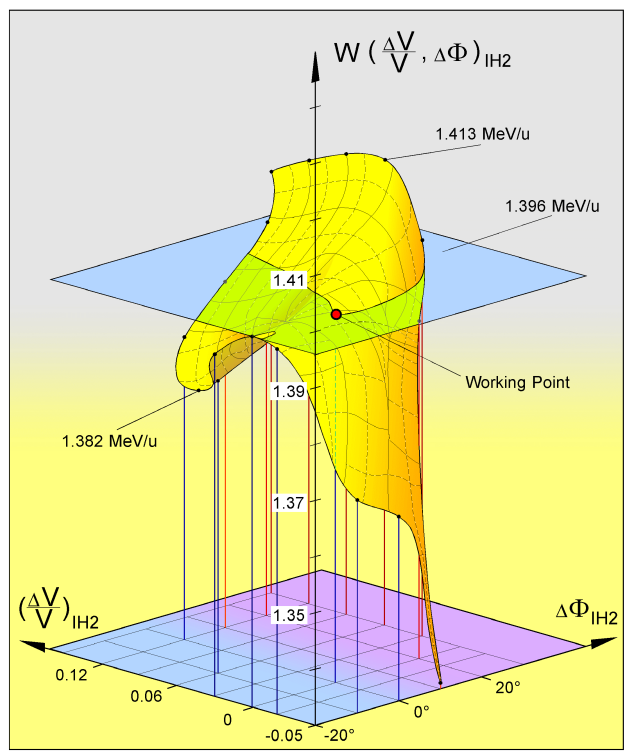

Fig. 2: Dependence of the IH2 exit energy on rf amplitude and phase of the same tank (LORASR simulation).

*Pulse Shaping by the LEBT Chopper:*
An important feature of the high current pulse shape is the rise time to full intensity: This defines the amount of beam energy to be handled, while machine parameters are changed or optimized. The linac beam pulse is defined by a chopper, which is located immediately in front of the RFQ. It has a rise time of about 0.5 µs [4]. After reaching the chopper voltage design levels pulse rise times of 5 µs are measured now as shown qualitatively by the dashed curves in Fig. 3, it is the time resolution of the current transformers. The curves measured during the commissioning show the quality of the flat top as well as the conservation of the beam pulse shape along the linac. This demonstrates the performance of the rf control loops, which are acting very well though the beam current is close to the design level of 10 mA $Ar^+$ in this example. In case of the IH-DTL, up to 44 % of the rf power are pumped into the beam. More detailed investigation on the pulse shape will be based on capacitive pick-up signals [5].

*Transmission:*

At a given beam current from the ion source the injected beam current into the linac was varied by cutting the horizontal beam emittance with slits. Especially the beam transmission along the RFQ depends very much on the slit width and is ranging from 60 % at $I_{in} = 18$ mA (measured in front of the quadruplet lense) up to around 100 % at $I_{in} < 4$ mA. The transmission along IH1 (IH2) is above 90 % (95 %) for the whole current range and is approaching 100 % at current levels below 2 mA. So far the RFQ has reached the design current (10 mA), while a maximum of 8 mA was measured behind of IH2.

*Transverse Emittance:*

Down to the RFQ exit the slit-grid emittance device was used for all current levels. At higher beam energies, high current and full pulse length, only single shot measurement techniques could be applied. Fig. 5 shows maximum current emittance plots from different commissioning steps. The measured LEBT emittance shows, that the matching into the RFQ causes beam losses at high input beam currents: the normalized RFQ acceptance was calculated to be 0.3 $\pi$ mm mrad only. At the exit of IH2 and at a beam current of 7.5 mA the measured norm. 80 % emittance area of. 0.45 $\pi$ mm mrad is well contained within the 90 % area resulting from LORASR calculations. More detailed studies about the RFQ transmission and comparisons with simulations are planned for the future. Bunch shape measurements with high resolution were done by using diamond detectors [10].

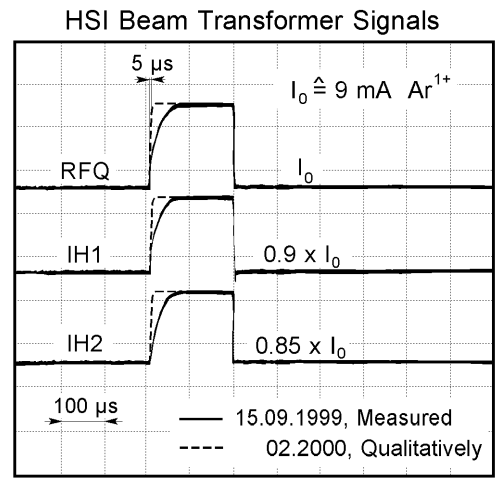

Fig. 3: Beam transformer signals along the HSI. Dashed lines correspond to an improved chopper performance (see Ref. 4)

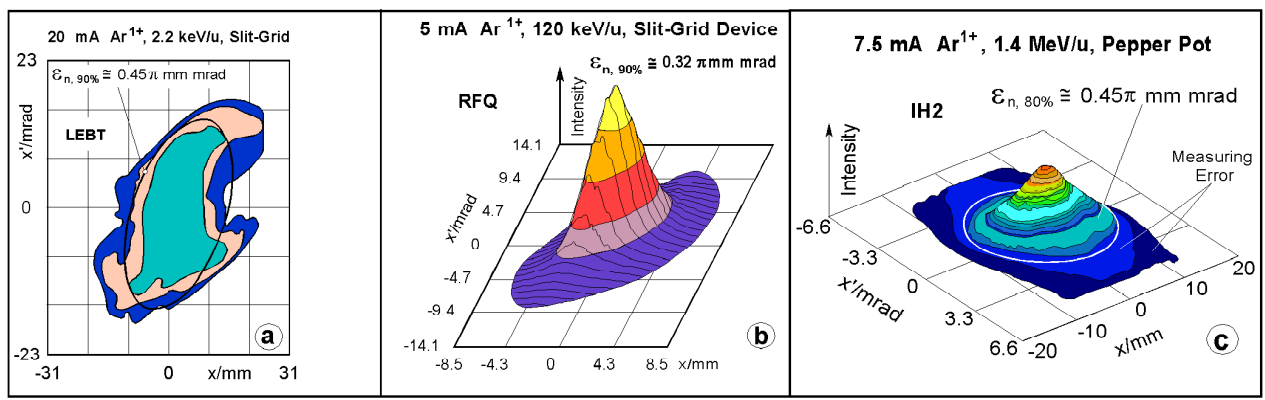

Fig. 4: Horizontal emittance measurements taken during different commissioning steps along the HSI